\documentstyle[12pt,worldsci]{article}
\input psfig.sty
\begin{document}
\setlength{\baselineskip}{2.6ex}
\def\nn{\noindent}
\def\Re{{\cal R \mskip-4mu \lower.1ex \hbox{\it e}\,}}
\def\Im{{\cal I \mskip-5mu \lower.1ex \hbox{\it m}\,}}
\def\ie{{\it i.e.}}
\def\eg{{\it e.g.}}
\def\etc{{\it etc}}
\def\etal{{\it et al.}}
\def\ibid{{\it ibid}.}
\def\sub#1{_{\lower.25ex\hbox{$\scriptstyle#1$}}}
\def\sul#1{_{\kern-.1em#1}}
\def\sll#1{_{\kern-.2em#1}}
\def\sbl#1{_{\kern-.1em\lower.25ex\hbox{$\scriptstyle#1$}}}
\def\ssb#1{_{\lower.25ex\hbox{$\scriptscriptstyle#1$}}}
\def\sbb#1{_{\lower.4ex\hbox{$\scriptstyle#1$}}}
\def\spr#1{^{\,#1}}
\def\spl#1{^{\!#1}}
\def\tev{\,{\rm TeV}}
\def\gev{\,{\rm GeV}}
\def\mev{\,{\rm MeV}}
\def\to{\rightarrow}
\def\rb{\ifmmode R_b\else $R_b$\fi}
\def\rc{\ifmmode R_c\else $R_c$\fi}
\def\ac{\ifmmode A_c\else $A_c$\fi}
\def\dmix{\ifmmode D^0-\bar D^0 \else $D^0-\bar D^0$\fi}
\def\dm{\ifmmode \Delta m_D \else $\Delta m_D$\fi}
\def\rb{\ifmmode R_b\else $R_b$\fi}
\def\mh{\ifmmode m\sbl H \else $m\sbl H$\fi}
\def\mch{\ifmmode m_{H^\pm} \else $m_{H^\pm}$\fi}
\def\mt{\ifmmode m_t\else $m_t$\fi}
\def\mc{\ifmmode m_c\else $m_c$\fi}
\def\mz{\ifmmode M_Z\else $M_Z$\fi}
\def\mw{\ifmmode M_W\else $M_W$\fi}
\def\mws{\ifmmode M_W^2 \else $M_W^2$\fi}
\def\mhs{\ifmmode m_H^2 \else $m_H^2$\fi}
\def\mzs{\ifmmode M_Z^2 \else $M_Z^2$\fi}
\def\mts{\ifmmode m_t^2 \else $m_t^2$\fi}
\def\mcs{\ifmmode m_c^2 \else $m_c^2$\fi}
\def\mchs{\ifmmode m_{H^\pm}^2 \else $m_{H^\pm}^2$\fi}
\def\ztwo{\ifmmode Z_2\else $Z_2$\fi}
\def\zone{\ifmmode Z_1\else $Z_1$\fi}
\def\mtwo{\ifmmode M_2\else $M_2$\fi}
\def\mone{\ifmmode M_1\else $M_1$\fi}
\def\bsg{\ifmmode b\to s\gamma\else $b\to s\gamma$\fi}
\def\tb{\ifmmode \tan\beta \else $\tan\beta$\fi}
\def\xw{\ifmmode x\sub w\else $x\sub w$\fi}
\def\ch{\ifmmode H^\pm \else $H^\pm$\fi}
\def\lum{\ifmmode {\cal L}\else ${\cal L}$\fi}
\def\inpb{\ifmmode {\rm pb}^{-1}\else ${\rm pb}^{-1}$\fi}
\def\infb{\ifmmode {\rm fb}^{-1}\else ${\rm fb}^{-1}$\fi}
\def\epem{\ifmmode e^+e^-\else $e^+e^-$\fi}
\def\ppb{\ifmmode \bar pp\else $\bar pp$\fi}
\def\subw{_{\rm w}}
\def\half{\textstyle{{1\over 2}}}
\def\elli{\ell^{i}}
\def\ellj{\ell^{j}}
\def\ellk{\ell^{k}}
\newskip\zatskip \zatskip=0pt plus0pt minus0pt
\def\matth{\mathsurround=0pt}
\def\lsim{\mathrel{\mathpalette\atversim<}}
\def\gsim{\mathrel{\mathpalette\atversim>}}
\def\atversim#1#2{\lower0.7ex\vbox{\baselineskip\zatskip\lineskip\zatskip
  \lineskiplimit 0pt\ialign{$\matth#1\hfil##\hfil$\crcr#2\crcr\sim\crcr}}}
\def\undertext#1{$\underline{\smash{\vphantom{y}\hbox{#1}}}$}

\rightline{\vbox{\halign{&#\hfil\cr
&SLAC-PUB-95-6821\cr
&April 1995\cr}}}
\title{{\bf SEARCHING FOR NEW PHYSICS WITH CHARM}
\footnote{\noindent{Work Supported by the Department of Energy,
Contract DE-AC03-76SF00515}}
\footnote{\noindent{Presented at {\it Lafex International School on High Energy
Physics (LISHEP95)}, Rio de Janeiro, Brazil, February 6-22, 1995}}
}
\author{J.L.\ HEWETT\\
\vspace{0.3cm}
{\em Stanford Linear Accelerator Center, Stanford University, Stanford, CA
94309, USA}}
\maketitle

\begin{center}
\parbox{13.0cm}
{\begin{center} ABSTRACT \end{center}
{\small\hspace*{0.3cm}
We consider the prospect of using the charm system as a
laboratory for probing new physics.
The theoretical expectations for rare charm meson decays, \dmix, and
charm quark asymmetries in $Z$ decays are examined in the Standard Model.
The effects of new physics from several classes of non-standard dynamical
models are summarized for these phenomena.
}}
\end{center}

\renewcommand{\thefootnote}{\arabic{footnote}}

\section{Overview}

One of the outstanding problems in particle physics
is the mysterious origin of the fermion mass and mixing spectrum.
One approach in addressing this question is to perform a detailed study of
the properties of all fermions.
While investigations of the $K$ and $B$ systems
have and will continue to play a central role in our quest to understand flavor
physics, in-depth examinations of the charm-quark sector have yet to
be performed, leaving a gap in our knowledge.  Since charm is the only heavy
charged $+2/3$ quark presently accessible to experiment, it provides the
sole window of opportunity to examine flavor physics in this sector.  In
addition, charm allows a complimentary probe of Standard Model (SM) physics
and beyond to that attainable from the down-quark sector.

Due to the effectiveness of the GIM mechanism, short distance SM contributions
to rare charm processes are very small.  Most reactions are thus dominated
by long range effects which are difficult to reliably calculate.  However,
for some interactions, there exists a window for the potential observation of
new physics.  In fact, it is precisely because the SM flavor changing neutral
current (FCNC) rates are so small that charm provides an untapped
opportunity to discover new effects and offers a
detailed test of the SM in the up-quark sector.  In this talk, we first review
the expectations for rare $D$ meson decays, focusing on radiative charm
decays.  We next discuss \dmix\ mixing, first in the SM, then in a variety of
models with new interactions.  We then finish with a summary of new physics
effects in charm quark asymmetries in $Z$ decays.

\section{Rare Decays of Charm}

FCNC decays of charm mesons include the processes $D^0\to\ell^+\ell^-,
\gamma\gamma$, and $D\to X_u+\gamma, X_u+\nu\bar\nu, X_u+\ell^+\ell^-$, with
$\ell=e, \mu$.  They proceed via electromagnetic or weak penguin
diagrams as well as receiving contributions from box diagrams in some cases.
The short distance SM contributions to these decays are quite small, as
the masses of the quarks which participate inside the loops (d, s, and b)
are tiny, resulting in a very effective GIM mechanism.  The calculation of the
short distance rates for these processes is
straightforward and the transition amplitudes and standard loop integrals,
which are categorized in Ref.\ 1 for rare $K$ decays, are easily
converted to the $D$ system.  The loop integrals relevant for
$D^0\to\gamma\gamma$ may be found in Ref.\ 2.  Employing the GIM
mechanism results in the general expression for the amplitudes,
\begin{equation}
{\cal A}\sim V_{cs}V^*_{us}[F(x_s)-F(x_d)]+V_{cb}V^*_{ub}[F(x_b)-F(x_d)] \,,
\end{equation}
with $V_{ij}$ representing the relevant Cabibbo-Kobayashi-Maskawa (CKM)
matrix elements, and $x_i\equiv m^2_i/M_W^2$.  The magnitude of the $s$- and
$b$-quark contributions are generally comparable
as the larger CKM factors compensate for the small strange quark mass.
The values of the resulting inclusive short distance branching fractions, are
summarized in Table 1, along with the current experimental
bounds\cite{pdg,raredk}.
\begin{table}
\centering
\begin{tabular}{|l|c|c|c|} \hline\hline
Decay Mode & Experimental Limit & $B_{S.D.}$ & $B_{L.D.}$ \\ \hline
$D^0\to\mu^+\mu^-$ & $<1.1\times 10^{-5}$ & $(1-20)\times 10^{-19}$ &
$<3\times 10^{-15}$ \\
$D^0\to e^+e^-$ & $<1.3\times 10^{-4}$ &  & \\
$D^0\to\mu^\pm e^\mp$ & $<1.0\times 10^{-4}$ & $0$ & $0$ \\ \hline
$D^0\to\gamma\gamma$ & --- & $10^{-16}$ & $<3\times 10^{-9}$ \\ \hline
$D\to X_u+\gamma$ & & $(4-8)\times 10^{-12}$ & \\
$D^0\to\rho^0\gamma$ & $<1.4\times 10^{-4}$ & & $(1-5)\times 10^{-6}$ \\
$D^0\to\phi^0\gamma$ & $<2.0\times 10^{-4}$ & & $(0.1-3.4)\times 10^{-5}$
\\ \hline
$D\to X_u+\ell^+\ell^-$ & & $4\times 10^{-9}$ & \\
$D^0\to\pi^0\mu\mu$ & $<1.7\times 10^{-4}$ & & \\
$D^0\to\bar K^0 ee/\mu\mu$ & $<17.0/2.5\times 10^{-4}$ & &
$<2\times 10^{-15}$ \\
$D^0\to\rho^0 ee/\mu\mu$ & $<2.4/4.5\times 10^{-4}$ & & \\
$D^+\to\pi^+ee/\mu\mu$ & $<250/4.6\times 10^{-5}$ & few$\times 10^{-10}$ &
$<10^{-8}$ \\
$D^+\to K^+ee/\mu\mu$ & $<480/8.5\times 10^{-5}$ & & $<10^{-15}$ \\
$D^+\to\rho^+\mu\mu$ & $<5.8\times 10^{-4}$ & & \\
\hline
$D^0\to X_u+\nu\bar\nu$ & & $2.0\times 10^{-15}$ & \\
$D^0\to\pi^0\nu\bar\nu$ & --- & $4.9\times 10^{-16}$ & $<6\times 10^{-16}$ \\
$D^0\to\bar K^0\nu\bar\nu$ & --- & & $<10^{-12}$ \\
$D^+\to X_u+\nu\bar\nu$ & --- & $4.5\times 10^{-15}$ & \\
$D^+\to\pi^+\nu\bar\nu$ & --- & $3.9\times 10^{-16}$ & $<8\times 10^{-16}$ \\
$D^+\to K^+\nu\bar\nu$ & --- & & $<10^{-14}$ \\ \hline\hline
\end{tabular}
\caption{Standard Model predictions for the branching fractions due to short
and long distance contributions for various rare $D$ meson decays. Also
shown are the current experimental limits.}
\end{table}
The corresponding short distance
exclusive rates are typically an order of magnitude less than the inclusive
case.  We note that the
transition $D^0\to\ell^+\ell^-,$ is helicity suppressed;
the range given for this branching fraction,
$(1-20)\times 10^{-19}$, indicates the effect of varying the parameters
in the ranges $f_D=0.15-0.25$ GeV and $m_s=0.15-0.40$ GeV.

The calculation of the long distance branching fractions are plagued with
hadronic uncertainties and the estimates listed in Table 1 convey
an upper limit on the size of these effects rather than an actual value.
These estimates have been computed by considering various intermediate
particle states (\eg, $\pi, K, \bar K, \eta, \eta', \pi\pi,\ {\rm or}\ K\bar
K$)
and inserting the known rates for the decay of the intermediate particles
into the final state of interest.  In all cases we see that the long
distance contributions overwhelm those from SM short distance physics

The radiative decays, $D\to X_u+\gamma$, warrant further discussion.  Before
QCD corrections are applied, the short distance inclusive rate is very
small, $B(c\to u\gamma)=1.4\times 10^{-17}$; however, the QCD corrections
greatly enhance this rate.  These corrections have recently been
calculated\cite{radcharm} employing
an operator product expansion, where the effective Hamiltonian
is evolved at leading logarithmic order from the electroweak scale down to
$\mu\sim m_c$ by the Renormalization Group Equations.  The evolution is
performed in two successive steps; first from the electroweak scale down to
$m_b$ working in an effective 5 flavor theory, and then to $\mu<m_b$ in an
effective 4 flavor theory.  We note that care must be taken in the operator
expansion in order to correctly account for the CKM structure of the
operators.  This procedure results\cite{radcharm} in
$B(c\to u\gamma)= (4.21-7.94)\times
10^{-12}$, where the lower(upper) value in the numerical range corresponds
to the scale $\mu=2m_c(m_c)$.  The effects of the QCD corrections
are dramatic, and the rate is almost entirely due to operator mixing.
The stability of this result can be tested once the complete next-to-leading
order corrections are known.

The long range effects in radiative charm meson decays have also been recently
examined in Ref.\ 5.  These effects can be separated into two
classes, (i) pole amplitudes, which correspond to the annihilation processes
$c\bar q_1\to q_2\bar q_3$ with the photon radiating from any of the quarks,
and (ii) vector meson dominance (VMD) contributions, which are described by
the underlying
process $c\to q_1\bar q_2 q$ followed by the conversion $\bar q_2 q\to\gamma$.
In the first class, either the D meson experiences weak mixing with the
particle intermediate states before photon emission occurs (denoted as
as type I transition), or the photon is emitted before weak mixing, \ie, the
final state meson is created via weak mixing (designated as type II).
The case of pseudoscalar intermediate states was
considered in the type I amplitudes, since
the pseudoscalar-photon-final state meson coupling can be phenomenologically
inferred from data.  In type II transitions, the $D\gamma D^*_n$
vertices have not yet been experimentally determined and thus one must
rely on theoretical modeling.  We note that both of these amplitudes are
parity conserving due to the electromagnetic transition.  In VMD reactions,
the amplitudes have been calculated using both (i) a phenomenological approach,
which utilizes available data on $D\to MV$ transitions, and (ii) the
theoretical model of Bauer, Stech, and Wirbel\cite{bsw}.  The expectations
for the transition amplitude in each case are presented in Table 2, as well as
the resulting range of predicted branching fractions for various exclusive
decay modes.  We see that there is a wide range of predictions, and that the
long range effects completely dominate over the short distance physics.
Observation of several of these exclusive decays,
would test the theoretical modeling,
and would enable the scaling of predictions to the B sector with greater
accuracy.  This would be important for the case of $B\to\rho\gamma$,
which suffers from long distance uncertainties\cite{sandip}, and from which
one would like to extract the CKM matrix element $V_{td}$.

\begin{table}
\centering
\begin{tabular}{|l|ccc|c|c|} \hline\hline
Mode & \multicolumn{3}{c|}{${\cal A}^{\rm pc}$} & ${\cal A}^{\rm pv}$
& $B(D\to M\gamma)$ ($10^{-5}$) \\ \hline
  & P-I & P-II & VMD & VMD & \\ \hline
$D^{+}_{s}\to\rho^+\gamma$ & $8.2$ & $-1.9$ & $\pm 3.2$ & $\pm 2.8$ & $6-38$\\
$D^0\to {\bar K}^{*0}\gamma$ & $5.6$ & $-5.9$ & $\pm 3.8$ & $\pm (5.1-6.8)$ &
$7-12$ \\
$D^{+}_{s}\to b_1^+\gamma$ & $7.2$ & --- & --- & --- & $\sim 6.3$\\
$D^{+}_{s}\to a_1^+\gamma$ & $1.2$ & --- & --- & --- & $\sim 0.2$\\
$D^{+}_{s}\to a_2^+\gamma$ & $2.1$ & --- & --- & --- & $\sim 0.1$\\
$D^+\to\rho^+\gamma$ & $1.3$ & $-0.4$ & $\pm 1.6$ & $\pm 1.9$ & $2-6$\\
$D^+\to b_1^+\gamma$ & $1.2$ & --- & --- & --- & $\sim 3.5$\\
$D^+\to a_1^+\gamma$ & $0.5$ & --- & --- & --- & $\sim 0.04$\\
$D^+\to a_2^+\gamma$ & $3.4$ & --- & --- & --- & $\sim 0.03$\\
$D^+_s\to K^{*+}\gamma$ & $2.8$ & $-0.5$ & $\pm 0.9$ & $\pm 1.0$ & $-.8-3$ \\
$D^+_s\to K_2^{*+}\gamma$ & $6.0$ & --- & --- & --- & $\sim 0.2$\\
$D^0\to\rho^0\gamma$ & $0.5$ & $-0.5$ & $\pm (0.2-1.0)$ & $\pm (0.6-1.0)$ &
$0.1-0.5$\\
$D^0\to\omega^0\gamma$ & $0.6$ & $-0.7$ & $\pm 0.6$ & $\pm 0.7$ & $\simeq
0.2$\\
$D^0\to\phi^0\gamma$ & $0.7$ & $-1.6$ & $\pm (0.6-3.5)$ & $\pm (0.9-2.1)$ &
$0.1-3.4$ \\
$D^+\to K^{*+}\gamma$ & $0.4$ & $-0.1$ & $\pm 0.4$ & $\pm 0.4$ & $0.1-0.3$\\
$D^0\to K^{*0}\gamma$ & $0.2$ & $-0.3$ & $\pm 0.2$ & $\pm 0.2$ & $\simeq
0.01$\\
\hline\hline
\end{tabular}
\caption{Predictions for the amplitudes (in units of $10^8$ GeV$^{-1}$) and
branching fractions of exclusive
charm decays due to long distance contributions.}
\end{table}

Lepton flavor violating decays, \eg, $D^0\to\mu^\pm e^\mp$ and
$D\to X+\mu^\pm e^\mp$, are strictly forbidden in the SM with massless
neutrinos.  In a model with massive non-degenerate neutrinos and
non-vanishing neutrino mixings, such as in four generation models,
$D^0\to\mu^\pm e^\mp$ would be mediated by box diagrams with the massive
neutrinos being exchanged internally.  LEP data restricts\cite{neutr} heavy
neutrino mixing with $e$ and $\mu$ to be $|U_{Ne}U^*_{N\mu}|^2<7\times
10^{-6}$ for a neutrino with mass $m_N>45$ GeV.  Consistency with this bound
constrains\cite{dipper} the branching fraction to be $B(D^0\to\mu^\pm
e^\mp)<6\times 10^{-22}$.  This result
also holds for a heavy singlet neutrino
which is not accompanied by a charged lepton.  The observation
of this decay at a larger rate than the above bound
would be a clear signal for the existence of
a different class of models with new physics.

Examining Table 1, we see that
the SM short distance contributions to rare charm decays are overwhelmed
by the long distance effects.  The observation of any of
these modes at a larger rate than what is predicted from long
distance interactions would provide a clear signal for new physics.
To demonstrate the magnitude of enhancements that are possible in
theories beyond the SM, we consider two examples (i) leptoquark exchange
mediating $D^0\to \mu^\pm e^\mp$ and (ii) a heavy $Q=-1/3$ quark
contributing to $c\to u\gamma$.  Leptoquarks are
color triplet particles which couple to a lepton-quark pair and are
naturally present in many theories beyond the SM which relate leptons and
quarks at a more fundamental level.  We parameterize their {\it a priori}
unknown couplings as $\lambda^2_{\ell q}/4\pi=F_{\ell q}\alpha$.  Leptoquarks
can mediate $D^0\to \mu^\pm e^\mp$ by tree-level exchange, however their
contributions are suppressed by angular momentum conservation.  From the
present limit $B(D^0\to \mu^\pm e^\mp)<10^{-4}$,
Davidson \etal\cite{sacha} derive the bound on the leptoquark mass $m_{lq}$
and coupling,
\begin{equation}
\sqrt{F_{eu}F_{\mu c}} < 4\times 10^{-3} {\alpha\over 4\pi}
\left[{m_{lq}\over 100{\rm GeV}}\right]^2 \,.
\end{equation}
These constraints surpass those from HERA\cite{hera}.
In the second example of new physics contributions, we examine a heavy $Q=-1/3$
quark, which may be present, \eg, as an iso-doublet fourth generation $b'$
quark, or as a singlet quark in $E_6$ grand unified theories.
The current bound\cite{pdg} on the mass of such an object
is $m_{b'}>85$ GeV, assuming that it decays via charged current interactions.
The heavy quark will then participate inside the penguin diagrams
which mediate $c\to u\gamma$, with the appropriate CKM factors.  From unitarity
considerations, the  fourth generation CKM factor will also contribute to
the coefficient of the current-current operator which dominates the branching
fraction via mixing.  The resulting $B(D\to X_u\gamma)$ is presented
in Fig. 1 as a function of the fourth generation CKM mixing factor,
for several values of the
heavy quark mass.  We see that a sizeable enhancement of the three generation
short distance rate is possible, however, the short distance rate is still
overpowered by the long range effects.

\begin{figure}[htbp]
\centerline{\psfig{figure=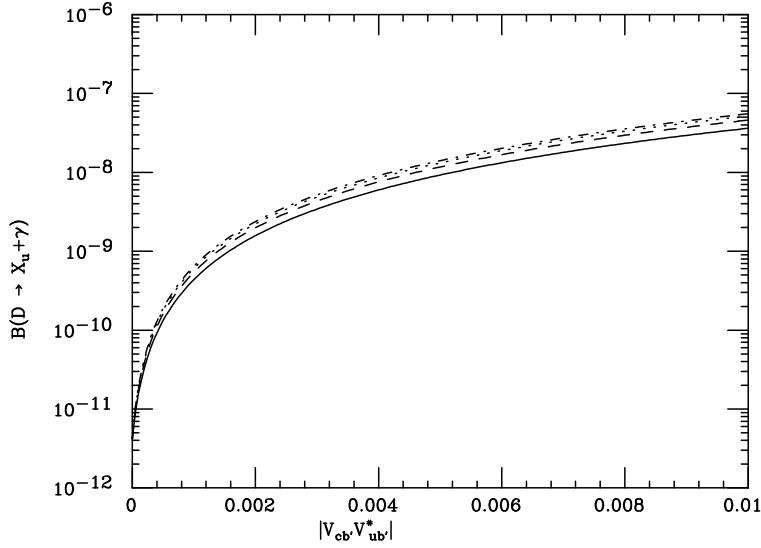,height=9cm,width=12cm,angle=90}}
\vspace*{-1cm}
\caption{\small Branching fraction for $D\to X_u\gamma$ in the four
generation SM as a function of the CKM mixing factor, with the solid, dashed,
dotted, dash-dotted curve corresponding to
$m_{b'}=100, 200, 300, 400$ GeV, respectively.}
\end{figure}

Non-SM contributions may also affect the purely leptonic decays
$D\to\ell\nu_\ell$.  Signatures for
new physics include the measurement of non-SM values for the absolute
branching ratios, or the observation of a deviation from the SM prediction
\begin{equation}
{B(D^+_{(s)}\to \mu^+\nu_\mu)\over B(D^+_{(s)}\to\tau^+\nu_\tau)}
={ {m^2_\mu\left( 1- m^2_\mu/ m^2_{D_{(s)}}\right)^2 }\over {m^2_\tau
\left( 1- m^2_\tau/ m^2_{D_{(s)}}\right)^2 }} \,.
\end{equation}
This ratio is sensitive to violations of $\mu-\tau$ universality.

As another example, we consider the case where the SM Higgs sector is
enlarged by an additional Higgs doublet.  These models generate
important contributions\cite{btaunu} to the decay $B\to\tau\nu_\tau$ and
it is instructive to examine their effects in the charm sector.  Two such
models, which naturally avoid tree-level flavor changing neutral currents,
are Model I, where one doublet ($\phi_2$) generates masses for all
fermions and the second ($\phi_1$) decouples from the fermion sector, and
Model II, where $\phi_2$ gives mass to the up-type quarks, while the down-type
quarks and charged leptons receive their mass from $\phi_1$.  Each doublet
receives a vacuum expectation value $v_i$, subject to the constraint that
$v_1^2+v_2^2=v^2_{\rm SM}$.  The charged Higgs boson present in these models
will mediate the leptonic decay through an effective four-Fermi interaction,
similar to that of the SM $W$-boson.  The $H^\pm$ interactions with
the fermion sector are governed by the Lagrangian
\begin{eqnarray}
{\cal L} & = & {g\over 2\sqrt 2 M_W}H^\pm[V_{ij}m_{u_i}A_u\bar
u_i(1-\gamma_5)d_j+V_{ij}m_{d_j}A_d\bar u_i(1+\gamma_5)d_j \nonumber \\
& & \quad\quad\quad\quad m_\ell A_\ell\bar\nu_\ell(1+\gamma_5)\ell]+h.c. \,,
\end{eqnarray}
with $A_u=\cot\beta$ in both models and $A_d=A_\ell=-\cot\beta(\tan\beta)$ in
Model I(II), where $\tan\beta\equiv v_2/v_1$.  In Models I and II, we obtain
the result
\begin{equation}
B(D^+\to\ell^+\nu_\ell)  =  B_{\rm SM}\left( 1+ {m_D^2\over m^2_{H^\pm}}
\right)^2 \,,
\end{equation}
where in Model II the $D_s^+$ decay receives an additional modification
\begin{equation}
B(D^+_s\to\ell^+\nu_\ell)  =  B_{\rm SM}\left[ 1+ {m_{D_s}^2\over m^2_{H^\pm}}
\left(1- \tan^2\beta{m_s\over m_c}\right)\right]^2 \,.
\end{equation}
In this case, we see that the effect of the $H^\pm$ exchange is independent
of the leptonic final state and the above prediction for the ratio in Eq. (3)
is unchanged.  This is because the $H^\pm$ contribution is proportional
to the charged lepton mass, which is then a common factor with the SM helicity
suppressed term.
However, the absolute branching fractions can be modified; this effect
is negligible in the decay $D^+\to\ell^+\nu_\ell$, but could be of
order a few percent in $D^+_s$ decay if $\tan\beta$ is very large.

\section{$D^0-\bar D^0$ Mixing}

Currently, the limits\cite{pdg} on \dmix\ mixing are from fixed target
experiments, with $x_D\equiv\Delta m_D/\Gamma<0.083$ (where $\Delta
m_D=m_2-m_1$
is the mass difference), yielding $\dm<1.3\times 10^{-13}$ GeV.
The bound on the ratio of wrong-sign to right-sign final
states is $r_D\equiv\Gamma(D^0\to\ell^-X)/\Gamma(D^0\to\ell^+X)<3.7\times
10^{-3}$, where
\begin{equation}
r_D\approx {1\over 2}\left[ \left( {\Delta m_D\over\Gamma}\right)^2 +
\left( {\Delta\Gamma\over 2\Gamma}\right)^2\right] \,,
\end{equation}
in the limit $\Delta m_D/\Gamma, \Delta\Gamma/\Gamma\ll 1$.  These analyses,
however, are based on the assumption that there is no interference between
the mixing signal and the dominant background of doubly Cabbibo suppressed
decays.  It has recently been noted\cite{yossi} that while this assumption
may be valid for the SM (since the expected size of the mixing is small), it
does not necessarily apply in models with new physics where \dmix\ mixing
is potentially large.

The short distance SM contributions to \dm\ proceed through a $W$ box diagram
with internal $d,s,b$-quarks.  In this case the external momentum, which is
of order $m_c$, is communicated to the light quarks in the loop and
can not be neglected.  The effective Hamiltonian is
\begin{equation}
{\cal H}^{\Delta c=2}_{eff} = {G_F\alpha\over 8\sqrt 2\pi x_w}\left[
|V_{cs}V^*_{us}|^2 \left(I_1^s {\cal O}-m_c^2I_2^s {\cal O'}\right)+
|V_{cb}V^*_{ub}|^2\left( I_3^b {\cal O}-m_c^2I_4^b {\cal O'}\right) \right] \,,
\end{equation}
where the $I_j^{q}$ represent integrals\cite{datta} that are functions of
$m_{q}^2/M_W^2$ and $m_{q}^2/m_c^2$, and ${\cal O}=[\bar u\gamma_\mu
(1-\gamma_5)c]^2$ is the usual mixing operator while ${\cal O'}=[
\bar u(1+\gamma_5)c]^2$ arises in the case of non-vanishing external
momentum.  The numerical value of the short distance contribution is
$\dm\sim 5\times 10^{-18}$ GeV (taking $f_D=200\mev$).  The long distance
contributions have been computed via two different techniques: (i) the
intermediate particle dispersive approach
(using current data on the intermediate states) yields\cite{gusto}
$\dm\sim 10^{-16}$ GeV, and (ii) heavy quark
effective theory which results\cite{hqet} in $\dm\sim 10^{-17}$ GeV.
Clearly, the SM predictions lie far below the
present experimental sensitivity!  We see that the gap between the short and
long distance expectations is not that large, and hence the opportunity exists
for new physics to reveal itself.

One reason the SM short distance
expectations for \dmix\ mixing are so small is that there
are no heavy particles participating in the box diagram to enhance the rate.
Hence the first extension to the SM that we consider is the
addition\cite{four} of a heavy $Q=-1/3$ quark.
We can now neglect the external momentum and \dm\ is given
by the usual expression\cite{inlim},
\begin{equation}
\dm={G_F^2M_W^2m_D\over 6\pi^2}f_D^2B_D|V_{cb'}V_{ub'}^*|^2F(m^2_{b'}/M_W^2)
\,.
\end{equation}
The value of \dm\ is displayed in this model in Fig.\ 2(a) as a function of the
overall CKM mixing factor for various values of the heavy quark mass.  We see
that \dm\ approaches the current experimental range for large values of the
mixing factor.

Next we examine two-Higgs-doublet models discussed above which
avoid tree-level FCNC by introducing a global symmetry.
The expression for \dm\ in these models can be found in Ref.\ 18.
{}From the Lagrangian in Eq. (4) it is clear that Model I will only modify the
SM
result for very small values of $\tan\beta$, and this region is already
excluded\cite{bhp,cleobsg} from existing data on
$b\to s\gamma$ and $B_d^0-\overline B_d^0$
mixing.  However, enhancements can occur in Model II for large values of
$\tan\beta$, as demonstrated in Fig.\ 2(b).

We now consider the case of extended Higgs sectors without natural flavor
conservation.  In these models the above requirement of a global symmetry
which restricts each fermion type to receive mass from only one doublet is
replaced\cite{fcnch} by approximate flavor symmetries which act on the
fermion sector.  The Yukawa couplings can then possess a structure which
reflects the observed fermion mass and mixing
hierarchy.  This allows the low-energy FCNC limits to be evaded as the
flavor changing couplings to the light fermions are small.  We employ the
Cheng-Sher ansatz\cite{fcnch}, where the flavor changing couplings of the
neutral Higgs are $\lambda_{h^0f_if_j}\approx (\sqrt 2G_F)^{1/2}
\sqrt{m_im_j}\Delta_{ij}$, with the $m_{i(j)}$ being the relevant fermion
masses and $\Delta_{ij}$ representing a combination of mixing angles.
$h^0$ can now contribute to \dm\ through tree-level exchange
as well as mediating \dmix\ mixing by $h^0$ and t-quark virtual
exchange in a box diagram.  These latter contributions only compete with those
from the tree-level process for large values of $\Delta_{ij}$.  In Fig. 2(c-d)
we show the value of \dm\ in this model from the tree-level and box diagram
contribution, respectively.

The last contribution to \dmix\ mixing that we will discuss here is that
of scalar leptoquark bosons.
They participate in \dm\ via virtual exchange inside a box
diagram\cite{sacha}, together with a charged lepton or neutrino.
Assuming that there is no leptoquark-GIM mechanism, and taking both exchanged
leptons to be the same type, we obtain the restriction
\begin{equation}
{F_{\ell c}F_{\ell u}\over m^2_{lq}} <
{196\pi^2 \dm\over (4\pi\alpha f_D)^2m_D} \,,
\end{equation}
where $F_{\ell q}$ is defined in the previous section.
The resulting constraints in the leptoquark coupling-mass plane are presented
in Fig. 2(e), assuming that a limit of $\dm<10^{-13}$ could be obtained
from experiment.

\nn
\begin{figure}[htbp]
\centerline{
\psfig{figure=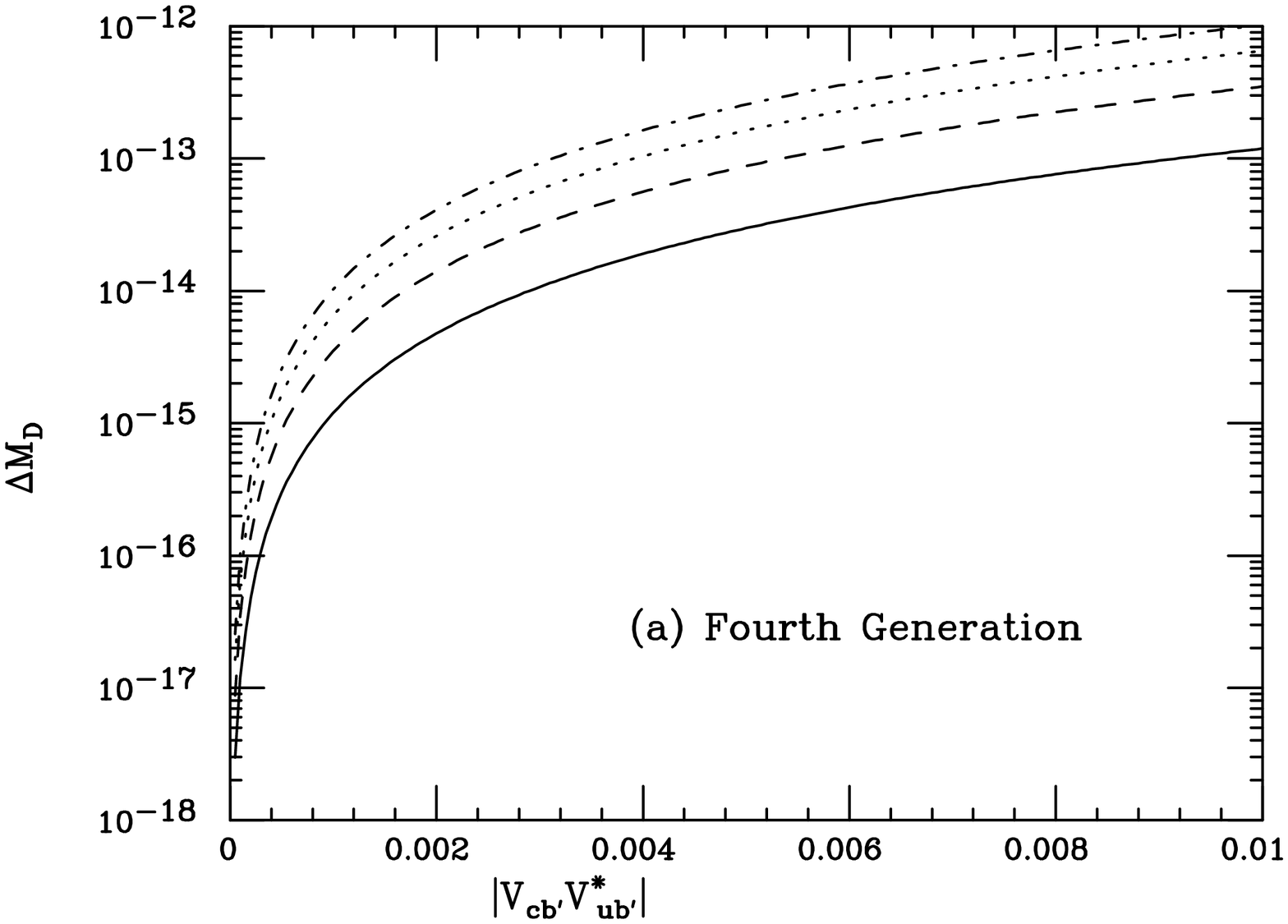,height=7.cm,width=8cm,angle=90}
\hspace*{-5mm}
\psfig{figure=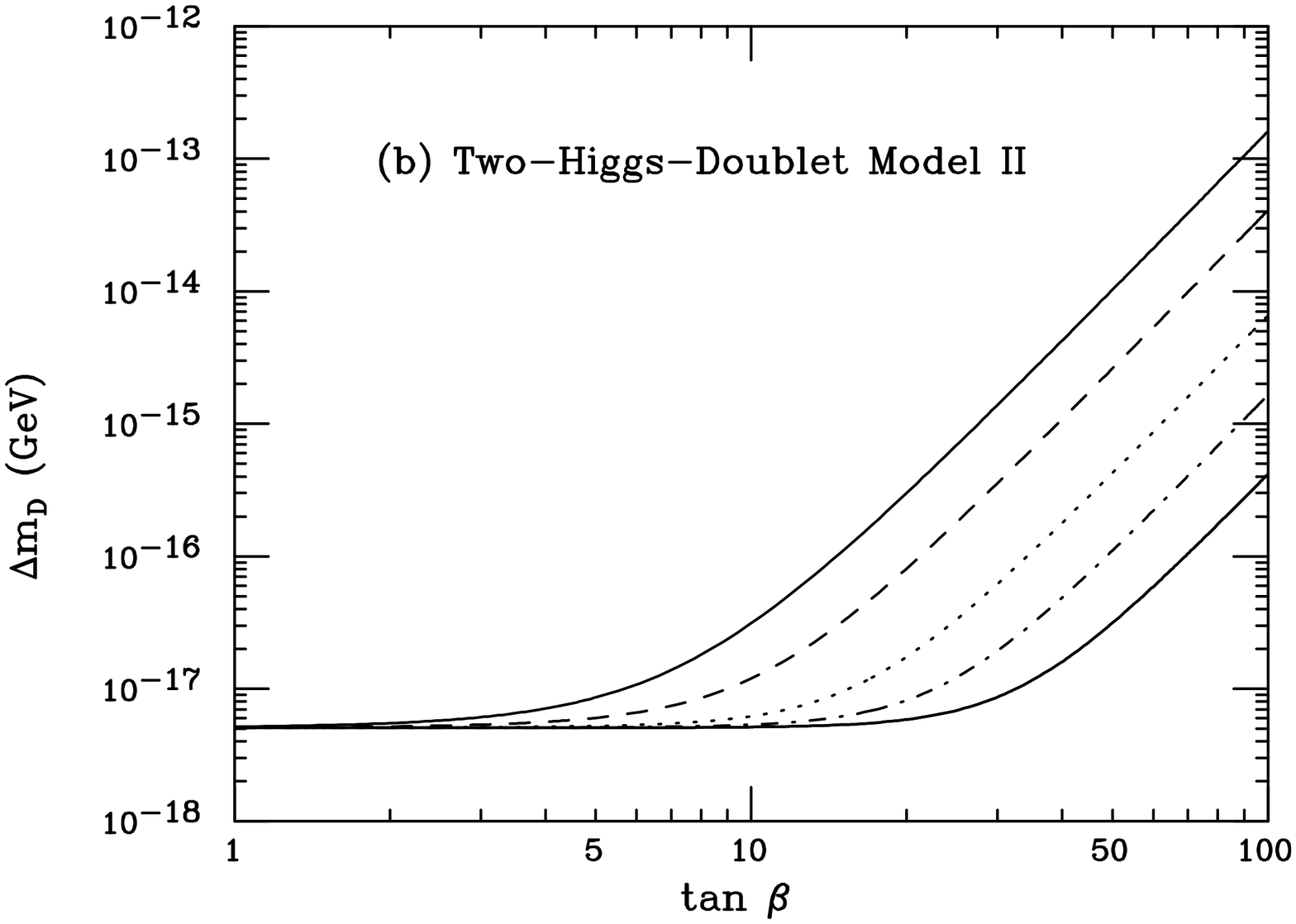,height=7.cm,width=8cm,angle=90}}
\vspace*{-0.75cm}
\centerline{
\psfig{figure=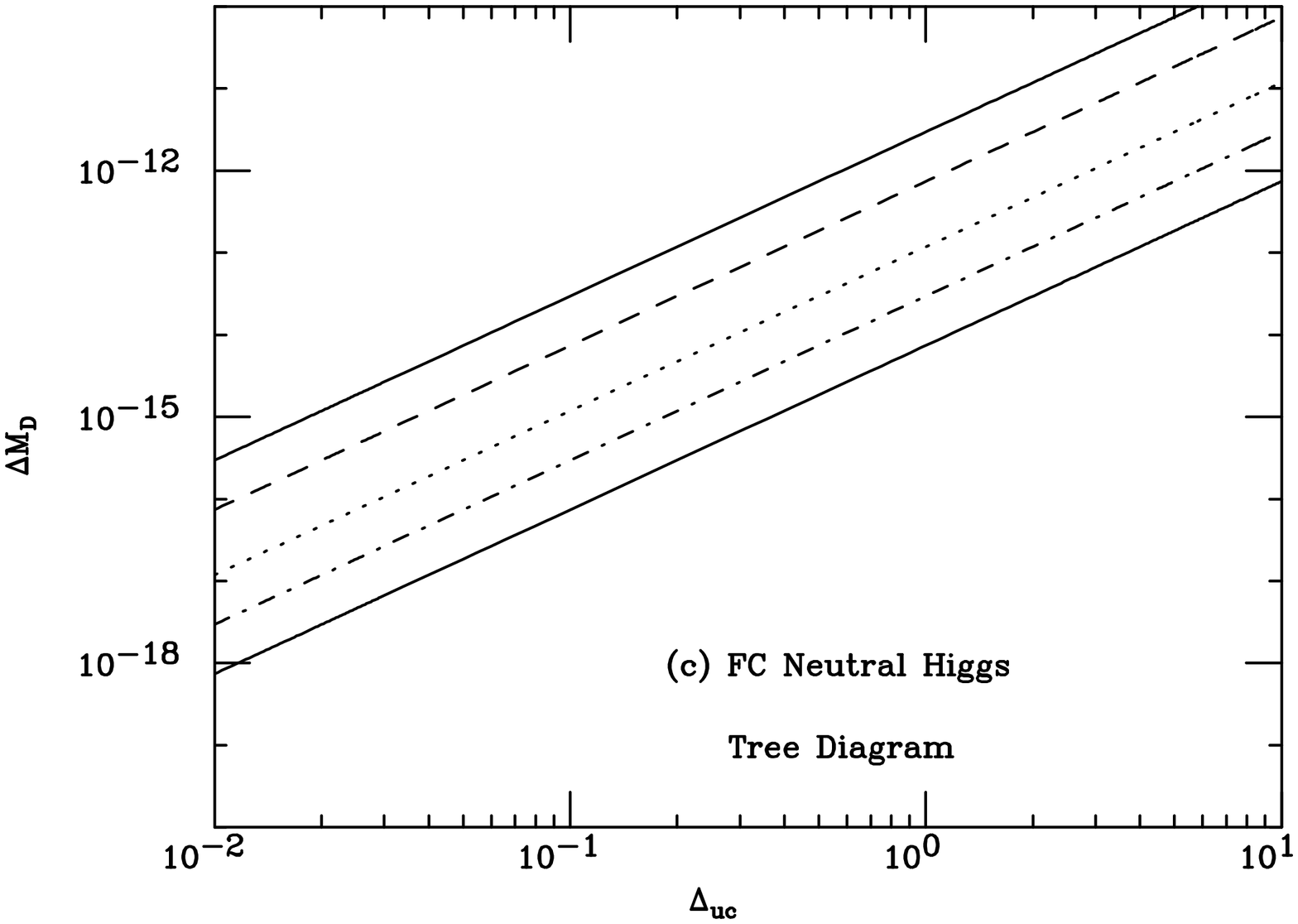,height=7.cm,width=8cm,angle=90}
\hspace*{-5mm}
\psfig{figure=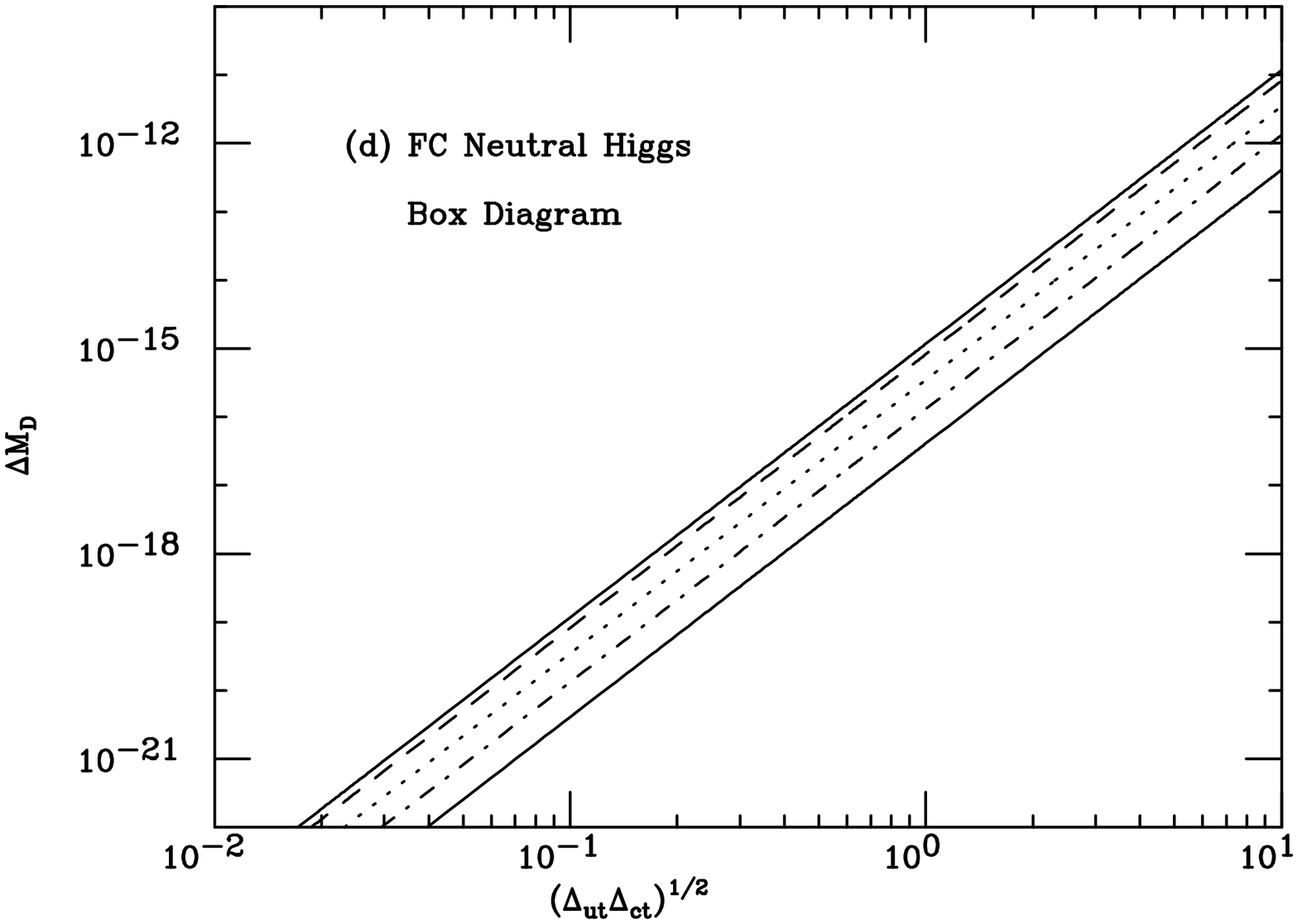,height=7.cm,width=8cm,angle=90}}
\vspace*{-0.75cm}
\centerline{
\psfig{figure=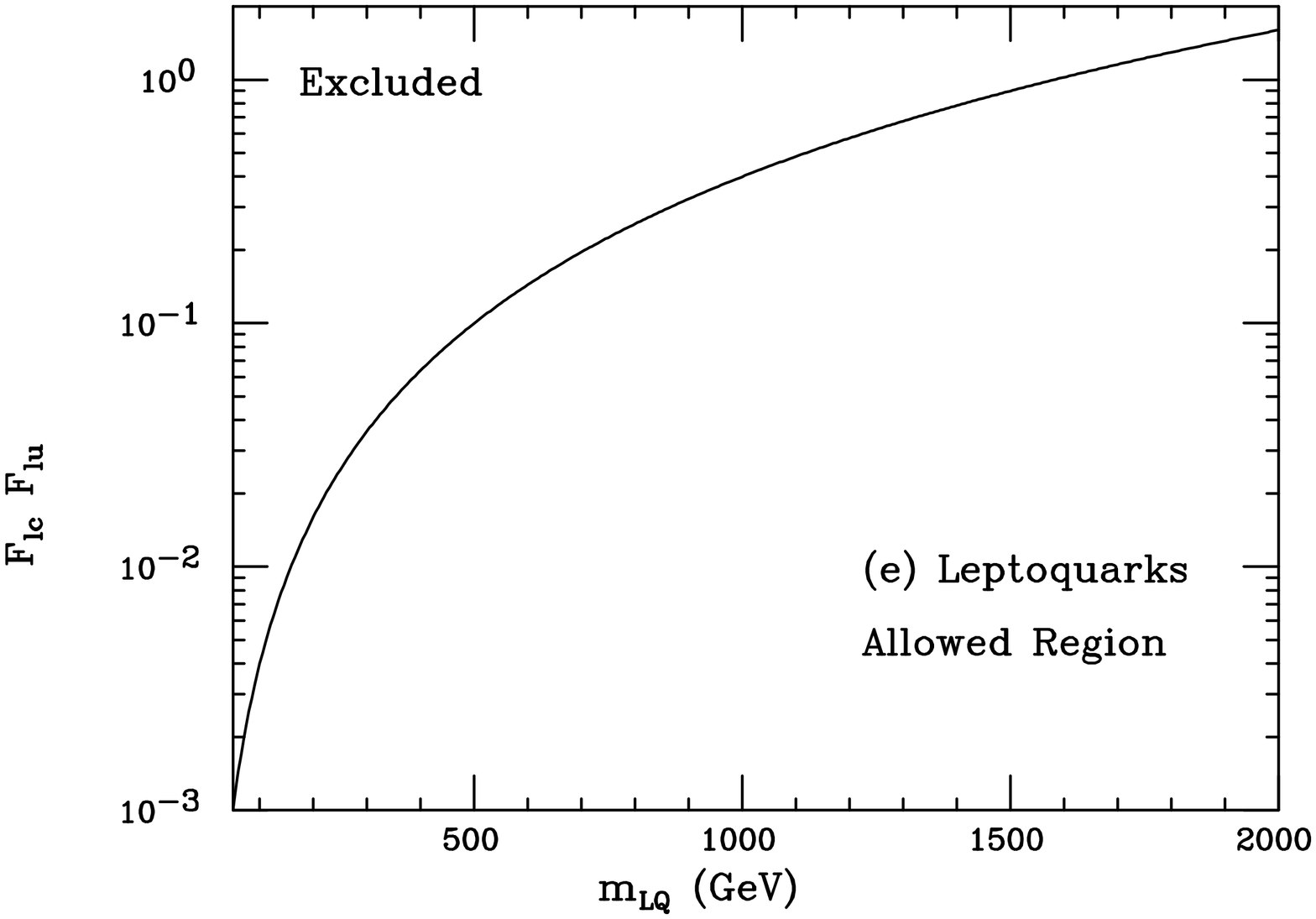,height=7.cm,width=10cm,angle=90}}
\vspace*{-1cm}
\caption{\small \dm\ in (a) the four generation SM with the same labeling as in
Fig. 1, (b) in two-Higgs-doublet model
II as a function of $\tan\beta$ with, from top to bottom, the solid, dashed,
dotted, dash-dotted, solid curve representing $m_{H^\pm}=50, 100, 250, 500,
1000$ GeV.  The solid horizontal line corresponds to the present experimental
limit.  (c) Tree-level and (d) box diagrams contributions to \dm\
in the flavor changing Higgs model described in the text as a function of
the mixing factor for $m_h=50, 100, 250, 500, 1000$ GeV corresponding to the
solid, dashed, dotted, dash-dotted, and solid curves from top to bottom.
(e) Constraints in the leptoquark coupling-mass plane from \dm.}
\end{figure}

\section{Charm Quark Asymmetries in $Z$ Decays}

The SM continues to provide an excellent description of
precision electroweak data\cite{schaile}, especially in the light of the
discovery of the top-quark\cite{cdfdo} in the mass range predicted by this
data.  The only hint of a potential discrepancy is a mere $(2-2.5)\sigma$
deviation from SM expectations for the quantity $\rb\equiv\Gamma(Z\to b\bar b)/
\Gamma(Z\to {\rm hadrons})$.  A global fit to all LEP data gives the
result\cite{schaile} $\rb=0.2204\pm 0.0020$.  In this fit, the value of
\rb\ is highly correlated to the value of the corresponding quantity $R_c$,
which is measured to be $R_c=0.1606\pm 0.0095$.  In contract to the b-quark
case, this is in reasonable
agreement with the SM expectation of $\rc=0.171$ (as defined by ZFITTER
4.9\cite{zfit} with $m_t=174$ GeV, $m_h=300$ GeV, and $\alpha_s=0.125$).
The asymmetry parameter, $\ac\equiv 2v_ca_c/(v_c^2+a_c^2)$,
is also measured\cite{schaile}  at LEP
via the charm-quark forward backward asymmetry,
$A_{FB}(c)=0.75 A_eA_c=0.0760\pm 0.0089$ and at the SLC via the left-right
forward-backward asymmetry $A^{LR}_{FB}(c)=0.75A_c$.  The SLD value for
\ac\ is\cite{schaile} $0.58\pm 0.14$, while the SM predicts\cite{zfit} 0.668.
In the SM, \rb\ is sensitive to additional vertex corrections involving the
top-quark, while the remaining electroweak and QCD radiative corrections
largely cancel in the ratio.  In contrast, \rc\ contains no such additional
SM vertex corrections.

The existence of anomalous couplings between the c-quark and
the $Z$ boson could cause a significant shift\cite{tgr} in the value of \rc.
The lowest dimensional non-renormalizable
operators which can be added to the SM take the form of either
electric or magnetic dipole form factors.  Defining $\kappa$ and $\tilde\kappa$
as the real parts of the magnetic and electric dipole form factors,
respectively, (evaluated at $q^2=M_Z^2$) the interaction Lagrangian is
\begin{equation}
{\cal L}={g\over 2c_w}\bar c\left[ \gamma_\mu(v_b-a_b\gamma_5) +
{i\over 2m_b}\sigma_{\mu\nu}q^\nu(\kappa_c^Z-i\tilde\kappa_c^Z\gamma_5)
\right] cZ^\mu \,.
\end{equation}
The influence of these couplings on \rc\ and \ac\
is presented in Fig. 3(a) from Rizzo\cite{tgr}, where the ratio of
these quantities calculated with the above Lagrangian to that of the SM
(as defined by ZFITTER\cite{zfit}) is displayed.  In this figure the solid
(dashed) curves represent the predictions when $\kappa_c^Z$
($\tilde\kappa_c^Z$)
is taken to be non-zero, with the diamonds representing unit steps of $0.002$
in these parameters.  The position of the data is also shown.

The extended Higgs models without natural flavor conservation discussed above
can also affect the $Zc\bar c$ vertex.  In this case, there is an extra
vertex correction due to the exchange of the neutral Higgs and the top-quark.
This correction takes the form $(ig/2c_w)(G_Fm_tm_c/8\sqrt 2\pi^2)
\Delta_{ct}^2\gamma_\mu(v_c\delta v_c-a_c\delta a_c\gamma_5)$, where
$\delta v_c$ and
$\delta a_c$ are given by the loop integrals.  The effect of this correction
on the asymmetry parameter is shown in Fig. 3(b), where we see that only
very large values of $\Delta_{ct}$ yield deviations from the SM.

Extended electroweak gauge sectors which contain an extra neutral gauge boson
can modify the fermion couplings of the SM $Z$.  These alterations in the
couplings arise due to (i) a shift in the
values of $v_f$ and $a_f$ due to $Z-Z'$ mixing, (ii) an overall factor
of $\sqrt\rho=M_Z^{SM}/M_{Z_1}$ due to the shift in the mass of the lightest
physical $Z_1$, from that predicted in the SM, and (iii) a shift in the
value of $x_w=sin^2\theta_w$ defined as $x_w(M_{Z_1})$ and not $x_W(M_Z^{SM})$.
The variation in the $Z\to c\bar c$ width and in \ac\ for the extended gauge
models based on $E_6$ and SO(10) grand unified theories\cite{esix} are
shown in Fig. 3(c-d), respectively.  In the $E_6$ case, the solid lines
correspond to fixed values of the $Z-Z'$ mixing angle $\phi$ for a 1 TeV
$Z'$, and the length of the lines represents the variation of the model
parameter $-90^\circ\leq\theta_{E_6}\leq 90^\circ$.  In the Left-Right
Symmetric Model (based on SO(10)), \ac\ is displayed as a function of the
ratio of right- to left-handed coupling strength, $\kappa\equiv g_R/g_L$,
for various values of the $Z'$ mass.

\nn
\begin{figure}[htbp]
\centerline{
\psfig{figure=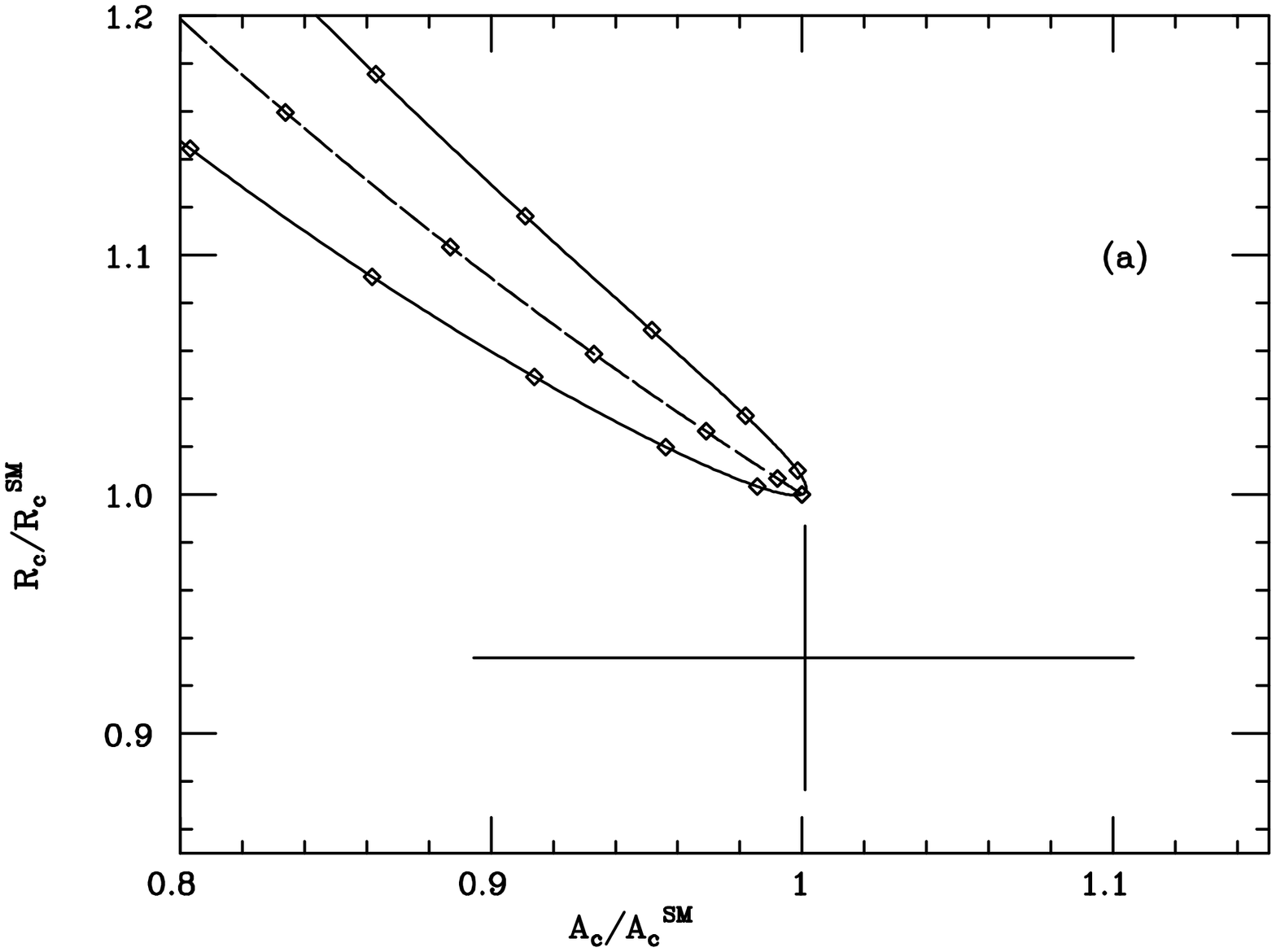,height=8.cm,width=8cm,angle=90}
\hspace*{-5mm}
\psfig{figure=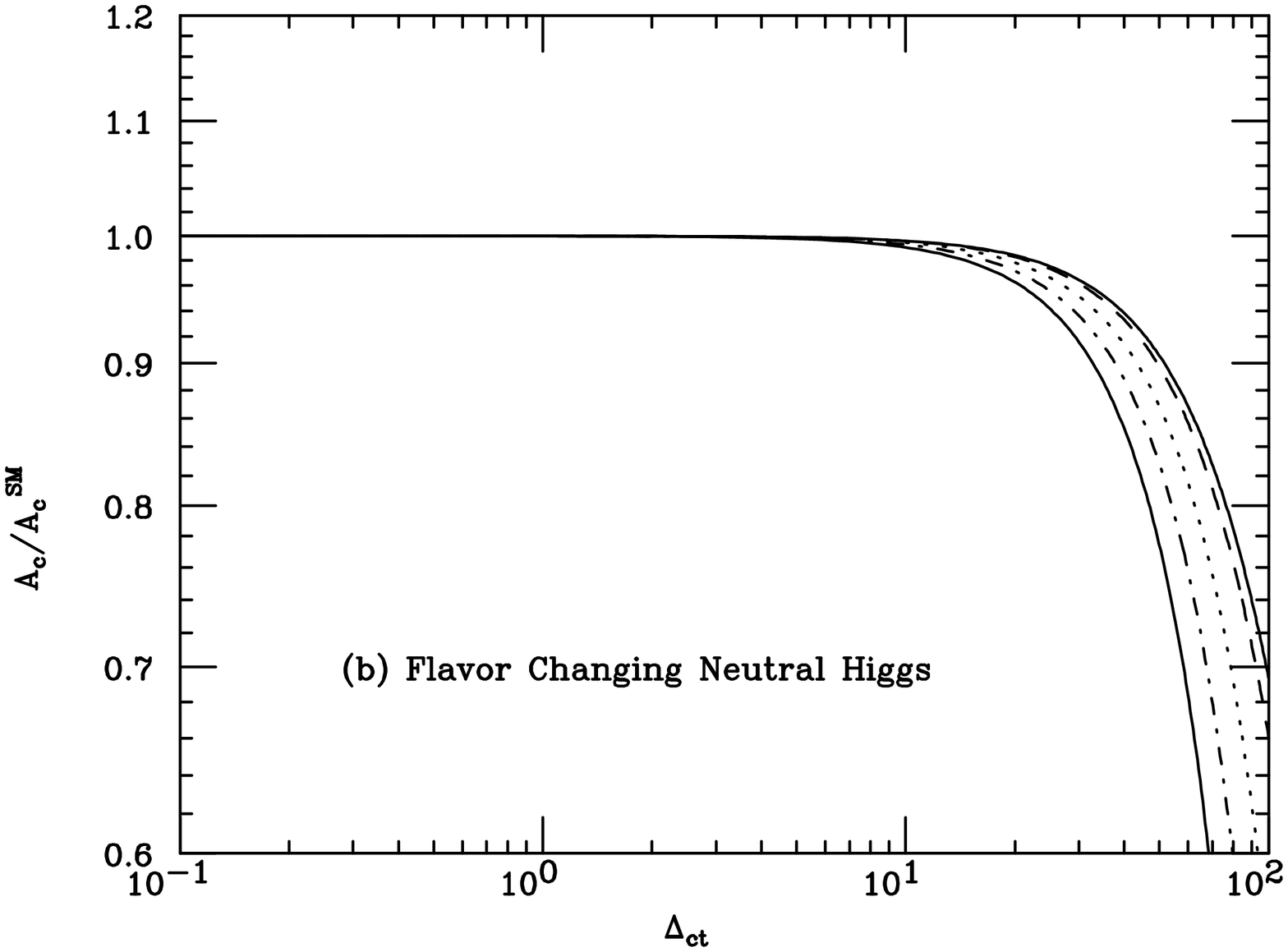,height=8.cm,width=8cm,angle=90}}
\vspace*{-0.75cm}
\centerline{
\psfig{figure=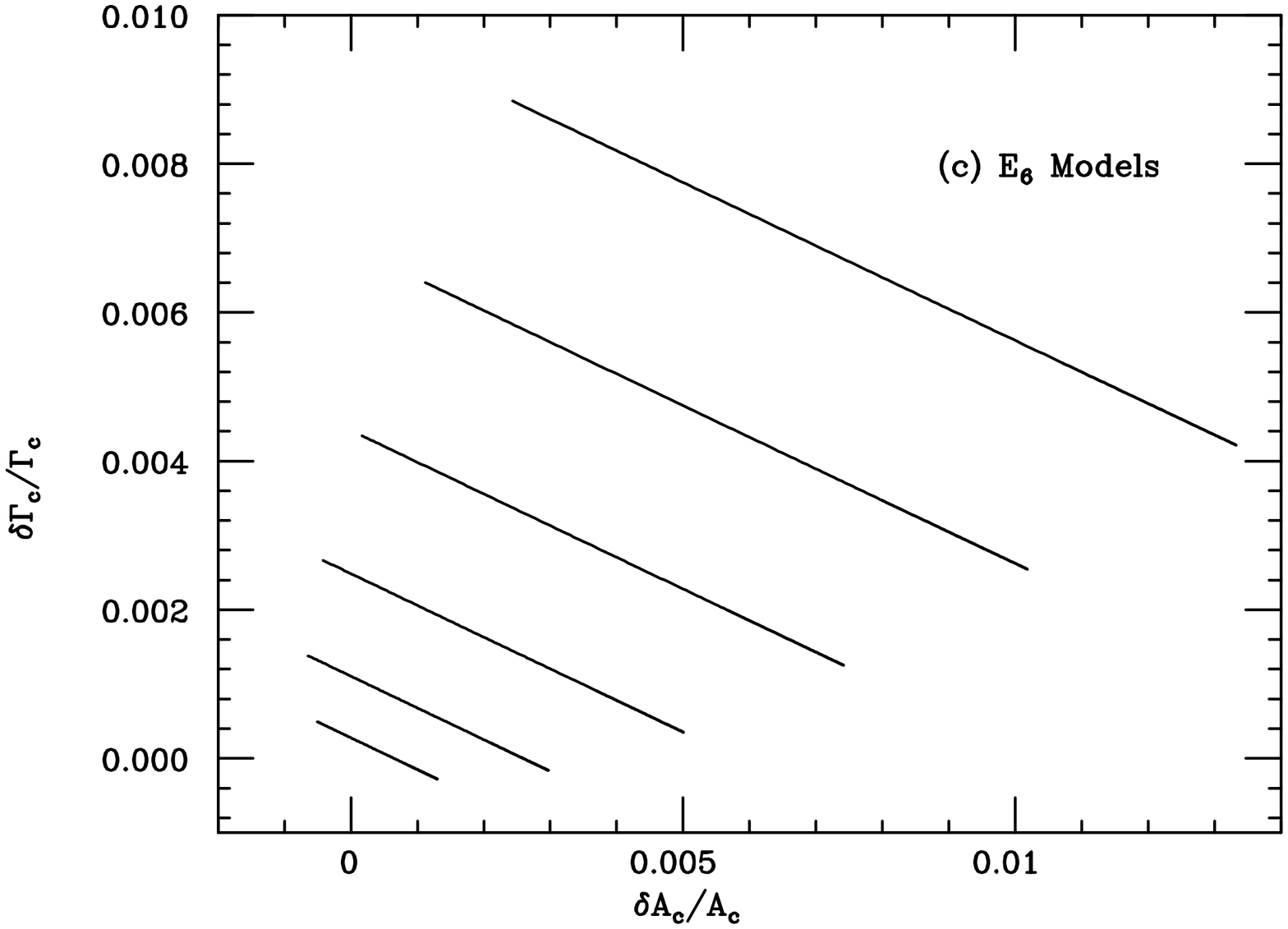,height=8.cm,width=8cm,angle=90}
\hspace*{-5mm}
\psfig{figure=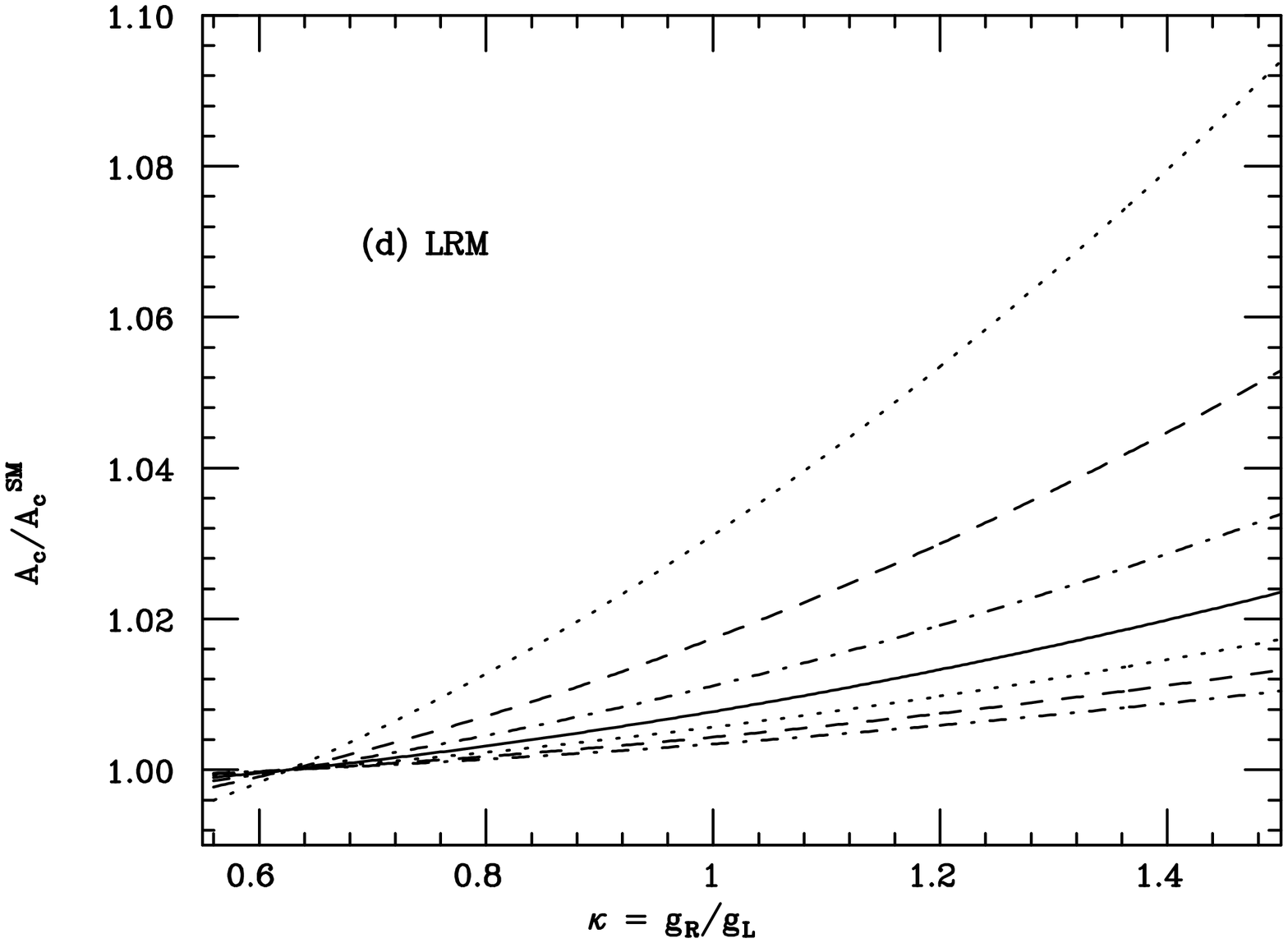,height=8.cm,width=8cm,angle=90}}
\vspace*{-1cm}
\caption{\small (a) The \rc\ and \ac\ plane, scaled to SM predictions, for
non-zero values of the electric
and magnetic dipole couplings from Rizzo in Ref. 24, where the diamonds
represent unit increments in these quantities in steps of 0.002.  The error
bars represent the data.  (b)  \ac, scaled to its SM value, in the flavor
changing neutral Higgs model as a function of the mixing parameter, for
$m_h=50,100,250,500,1000$ GeV with the same labeling as in Fig. 2(b).
(c)  Variations in the \ac\ - $Zc\bar c$ width plane in $E_6$ models, for
$|\phi|=0.006, 0.005, 0.004, 0.003, 0.002, 0.001$ from top to bottom with
a 1 TeV $Z'$.  (d) \ac\ in the Left-Right Symmetric Model as a function
of $\kappa\equiv g_R/g_L$, with a $Z'$ mass of 0.75, 1.0, 1.25, 1.5, 1.75,
2.0, 2.25
TeV corresponding to the dotted, dashed, dashdotted, solid, dotted dashed,
dashdotted curves from top to bottom.}
\end{figure}

\section{Summary}

In summary we see that there is a wide physics potential to motivate an
in-depth study of the charm system.  We urge our experimental colleagues
to study charm with the same precision that has and will be
achieved in the down-quark sector.
\vspace{1.0cm}

{\noindent\bf Acknowledgements}

I thank the organizers for providing the opportunity for me to attend this
stimulating workshop.
I am indebted to my collaborators G.\ Burdman, E.\ Golowich, and S.\ Pakvasa,
and I thank T. Rizzo for useful discussions.
\vspace{1.0cm}
%
\def\MPL #1 #2 #3 {Mod.~Phys.~Lett.~{\bf#1},\ #2 (#3)}
\def\NPB #1 #2 #3 {Nucl.~Phys.~{\bf#1},\ #2 (#3)}
\def\PLB #1 #2 #3 {Phys.~Lett.~{\bf#1},\ #2 (#3)}
\def\PR #1 #2 #3 {Phys.~Rep.~{\bf#1},\ #2 (#3)}
\def\PRD #1 #2 #3 {Phys.~Rev.~{\bf#1},\ #2 (#3)}
\def\PRL #1 #2 #3 {Phys.~Rev.~Lett.~{\bf#1},\ #2 (#3)}
\def\RMP #1 #2 #3 {Rev.~Mod.~Phys.~{\bf#1},\ #2 (#3)}
\def\ZP #1 #2 #3 {Z.~Phys.~{\bf#1},\ #2 (#3)}
\def\IJMP #1 #2 #3 {Int.~J.~Mod.~Phys.~{\bf#1},\ #2 (#3)}
\bibliographystyle{unsrt}

\begin{thebibliography}{99}
%
\bibitem{inlim}
T.\ Inami and C.S.\ Lim, Prog.\ Theor.\ Phys.\ {\bf 65}, 297 (1981).
%
\bibitem{mapram}
E.\ Ma and A.\ Pramudita, \PRD D24 2476 1981 .
%
\bibitem{pdg}
L.\ Montanet \etal, Particle Data Group, \PRD D50 1173 1994 .
%
\bibitem{raredk}
M.\ Selen, CLEO Collaboration, talk presented at {\em APS Spring Meeting},
Washington D.C., April 1994;
J.\ Cumalet, talk presented at {\it The Tau-Charm Factory in the Era of
B Factories and CESR}, Stanford, CA, August 1994.
%
\bibitem{radcharm}
G.\ Burdman, E.\ Golowich, J.L.\ Hewett, and S.\ Pakvasa, SLAC Report
SLAC-PUB-6692 (1994).
%
\bibitem{bsw}
M.\ Bauer, B.\ Stech, and M.\ Wirbel, \ZP C29 637 1985 , \ibid {\bf C34},
101 (1987).
%
\bibitem{sandip}
E.\ Golowich and S.\ Pakvasa, \PRD D51 1215 1995 ;
D.\ Atwood, B.\ Blok, and A.\ Soni, SLAC Report SLAC-PUB-6635 (1994) ;
G.\ Eilam, A. Ioannissian, and R.R. Mendel, Technion Report PH-95-4 (1995).
%
\bibitem{neutr}
A.\ Acker and S.\ Pakvasa, \MPL A7 1219 1992 .
%
\bibitem{dipper}
S.\ Pakvasa, in {\it CHARM2000 Workshop}, Fermilab, June 1994.
%
\bibitem{sacha}
S.\ Davidson, D.\ Bailey, and B.A.\ Campbell, \ZP C61 613 1994 .
%
\bibitem{hera}
M. Derrick \etal, (ZEUS Collaboration), DESY Report 94-07 (1994).
%
\bibitem{btaunu}
W.-S.\ Hou, \PRD D48 2342 1993 ; P.\ Krawczyk and S.\ Pokorski,
\PRL 60 182 1988 .
%
\bibitem{yossi}
G. Blaylock, A. Seiden, and Y. Nir, Univ. of California Santa Cruz
Report SCIPP 95/16 (1995); T. Liu, in
{\it CHARM2000 Workshop}, Fermilab, June 1994.
%
\bibitem{datta}
A.\ Datta, \ZP C27 515 1985 .
%
\bibitem{gusto} G.\ Burdman in {\it CHARM2000 Workshop}, Fermilab, June 1994;
J.\ Donoghue \etal, \PRD D33 179 1986 .
%
\bibitem{hqet}
H.\ Georgi, \PLB B297 353 1992 ; T.\ Ohl \etal, \NPB B403 605 1993 .
%
\bibitem{four}
K.S.\ Babu \etal, \PLB B205 540 1988 ; T.G.\ Rizzo, \IJMP A4 5401 1989 .
%
\bibitem{bhp}
J.L.\ Hewett, \PRL 70 1045 1993 ;
V.\ Barger, J.L.\ Hewett, and R.J.N.\ Phillips, \PRD D41 3421 1990 .
%
\bibitem{cleobsg}
R.\ Ammar \etal, CLEO Collaboration, \PRL 71 674 1993 ;
%
\bibitem{fcnch}
S.\ Pakvasa and H.\ Sugawara, \PLB 73B 61 1978 ;
T.P.\ Cheng and M.\ Sher,
\PRD D35 3484 1987 ; L.\ Hall and S.\ Weinberg, \PRD D48 979 1993 .
%
\bibitem{schaile}
D.\ Schaile, in {\it 27th International Conference on High Energy Physics},
Glasgow, Scotland, July 1994; U. Uwer in {\it 30th Rencontres de Moriond:
Electroweak Interactions and Unified Theories}, Meribel les Allures, France,
March 1995; H. Neal (SLD Collaboration), \ibid.
%
\bibitem{cdfdo}
F. Abe \etal, (CDF Collaboration), FERMILAB-PUB-95-022-E (1995);
S. Abachi \etal, (D0 Collaboration), FERMILAB-PUB-95-028-E (1995).
%
\bibitem{zfit}
The ZFITTER package: D.\ Bardin \etal, \ZP C44 493 1989 ; \NPB B351 1 1991 ;
\PLB B255 290 1991 ; CERN Report, CERN-TH-6443/92, 1992.
%
\bibitem{tgr}
T.G. Rizzo, \PRD D51 3811 1995 ; G. Kopp \etal, \ZP C65 545 1995 .
%
\bibitem{esix}
For a review and original references, see,
J.L. Hewett and T.G. Rizzo, \PR 183 193 1989 ; R.N. Mohapatra, {\it Unification
and Supersymmetry}, (Springer, New York, 1986).
%
\end{thebibliography}

\end{document}